\documentclass[journal=jpccck,manuscript=article,layout=twocolumn]{achemso}

\setkeys{acs}{abbreviations=true, articletitle=true,
biblabel=plain,
chaptertitle=true,
doi=true,
email=true,
etalmode=truncate,
keywords=true,
maxauthors=100,
super=true
}

\setcitestyle{super,open={},close={}}

\usepackage[utf8]{inputenc} 
\usepackage{textcomp} 
\usepackage[T1]{fontenc} 
\usepackage[portuguese,english]{babel} 
\usepackage[version=4]{mhchem} 
\usepackage{siunitx} 
\usepackage{graphicx} 
\graphicspath{{Figures/}} 
\usepackage{geometry} 
\usepackage[format=plain,justification=justified,singlelinecheck=false,font={normalsize,bf},labelfont=bf,labelsep=space]{caption} 
\usepackage{float} 
\usepackage{natbib} 
\usepackage{setspace} 
\usepackage{xkeyval} 
\usepackage{array} 
\usepackage{booktabs} 
\usepackage{listings} 
\usepackage{lmodern} 
\usepackage{mathpazo} 
\usepackage[breaklinks,colorlinks=true,allcolors=blue]{hyperref}
\hypersetup{colorlinks=true,citecolor=blue,urlcolor=blue}
\newcommand{\doi}[1]{\href{http://dx.doi.org/#1}{\nolinkurl{#1}}}

\usepackage{subfigure} 
\usepackage[compact]{titlesec} 
\usepackage{natmove} 
\usepackage{multirow} 
\usepackage{lipsum} 
\usepackage{calc} 
\usepackage{longtable} 
\usepackage{tabularx} 
\usepackage{placeins} 
\usepackage{tablefootnote}
\usepackage{adjustbox}
\usepackage{titletoc}
\usepackage{latexsym}
\usepackage{cancel}
\usepackage{amsmath}
\usepackage{amssymb}
\usepackage{amsfonts}
\usepackage{amstext}
\usepackage{float}

\DeclareSIUnit\angstrom{\text {Å}}
\DeclareSIUnit\bar{bar}

\usepackage[none]{hyphenat}
\usepackage{microtype}
\makeatletter
\let\l@addto@macro\relax
\makeatother
\usepackage[fontsize=12pt]{scrextend}

\SectionsOn
\SectionNumbersOn
\AbstractOn

\title[Title]{Raman Spectra and Excitonic Effects of the novel \ce{Ta2Ni3Te5} Monolayer}

\author{Alexandre C. Dias}
\affiliation[CIF]{University of Bras{\'{i}}lia, Institute of Physics and International Center of Physics, Bras{\'{i}}lia $70919$-$970$, DF, Brazil}
\email{alexandre.dias@unb.br}

\author{Raphael M. Tromer}
\affiliation{School of Engineering, Mackenzie Presbyterian University, Rua da Consola\c c\~ao, 930. S\~ao Paulo - SP. 01302-907. Brazil}

\author{Humberto R. Gutiérrez}
\affiliation{Department of Physics, University
of South Florida, Tampa, Florida 33620, United States}

\author{Douglas S. Galvão}
\affiliation[Unicamp]{State University of Campinas and Center for Computational Engineering and Sciences, Campinas $13083-970$, SP, Brazil}

\author{Elie A. Moujaes}
\affiliation[UNIR]{Physics Department, Federal University of Rondônia, $76801-974$, Porto Velho, Brazil}
\email{eamoujaes@unir.br}

\begin{document}

\maketitle

\begin{abstract}
We have investigated the Raman spectrum and excitonic effects of the novel two-dimensional \ce{Ta2Ni3Te5} structure. The monolayer is an indirect band gap semiconductor with an electronic band gap value of \SI{0.09}{\electronvolt}  and \SI{0.38}{\electronvolt}, determined using GGA-PBE and HSE06 exchange-correlation functionals, respectively. Since this structure is energetically, dynamically, and mechanically stable, it could be synthesized as a free-standing material. We identify ten Raman and ten infrared active modes for various laser energies, including those commonly used in Raman spectroscopy experiments. It was also observed that the contribution of \ce{Ni} atoms is minimal in most Raman vibrational modes. In contrast, most infrared vibrational modes do not involve the vibration of the \ce{Ta} atoms. As far as the optical properties are concerned, this monolayer shows a robust linear anisotropy, an exciton binding energy of \SI{287}{\milli\electronvolt}, and also presents a high reflectivity in the ultraviolet region, which is more intense for linear light polarization along the $x$ direction. 
\end{abstract}

\section{Introduction}

Graphene's experimental realization in the early 2000s \cite{Novoselov_666_2004, Geim_183_2007} marked a new frontier in solid-state physics and materials science. Graphene exhibits unique electronic and mechanical properties.\cite{Wang_699_2012} Graphene also renewed the interest in other new 2D materials, some of which possess interesting properties \cite{Correa_195422_2020, Lu_5204_2022, moujaes2019thermoelectric, moujaes2021optical} and functionalities,\cite{Dias_075202_2018, Dias_3265_2021, Dias_054001_2022}. Examples include the combination of different monolayers to form van der Waals (vdW) heterojunctions, allowing to tune a wide variety of properties \cite{Besse_041002_2021, Silveira_1671_2021, Silveira_9173_2022}. Unlike some bulk systems, the quantum confinement in those materials makes the excitonic quasi-particle effects relevant for a reliable characterization of their linear optical response,\cite{Dias_3265_2021, Moujaes_111573_2023} requiring sophisticated experimental and theoretical approaches.

Recently, vdW layered materials of the \ce{A2M_{1,3}X5} (\ce{A}=\ce{Ta,Nb}; \ce{M}=\ce{Pd,Ni} and \ce{X}=\ce{Se, Te}) family have received attention due to exotic properties, such as the quantum spin Hall effect in the \ce{Ta2Pd3Te5} monolayer,\cite{Guo_115145_2021, Wang_241408_2021} excitons in \ce{Ta2NiSe5},\cite{Wakisaka_02602_2009, Lu_14408_2017} and superconductivity in \ce{Nb2Pd3Te5} and doped \ce{Ta2Pd3Te5} structures.\cite{Higashihara_063705_2021}  Both \ce{Ta2Pd3Te5} and \ce{Ta2Ni3Te5}, which have been experimentally created in their layered bulk form,\cite{Zhang_011047_2024, Harrison_4811_2024}, exhibit intriguing topological properties.\cite{Guo_87_2022,Zhang_011047_2024} 

\ce{Ta2Pd3Te5} and \ce{Ta2Ni3Te5} share the same crystal structure,\cite{Jiang_067402_2024} consisting of rhombus-like \ce{Ta2Pd2}(\ce{Ta2Ni2}) clusters that act as building blocks within each layer. These clusters are linked into 1D chains, then loosely connected to form a 2D layer.\cite{Tan_2201_2024} Despite the structural similarities, both compounds exhibit distinct properties due to differences in the band gap of their normal states;\cite{Jiang_067402_2024} \ce{Ta2Pd3Te5} is classified as an excitonic insulator\cite{Jiang_067402_2024, Yao_097101_2024} whereas \ce{Ta2Ni3Te5} is a small band gap semiconductor, with an electronic band gap of \SI{31}{\milli\electronvolt} in its bulk form.\cite{Zhang_011047_2024,Jiang_067402_2024}

In particular, \ce{Ta2Ni3Te5} exhibits topological phase transitions under pressure and is also predicted to undergo a similar transition under strain \cite{Guo_115145_2021,Yang_020503_2023}, showing potential for various practical applications. In addition, \ce{Ta2Ni3Te5} also presents an in-plane anisotropy, which originates from its anisotropic crystal lattice and point group symmetry;\cite{Harrison_4811_2024}; this leads to a dependency of its electronic, optical, and thermal properties on the crystal orientation, hence enabling its use in anisotropic photoelectric and thermoelectric devices, similar to black phosphorus. \cite{Fei_6393_2014,Luo_9572_2015, Harrison_4811_2024} In fact, Harrison and co-workers investigated this in-plane anisotropy through polarized Raman spectroscopy, establishing a clear correlation between the structural and optical in-plane anisotropies in exfoliated few-layer \ce{Ta2Ni3Te5}.\cite{Harrison_4811_2024} 

In this work, we systematically investigated the electronic, optical, excitonic, and vibrational properties of the \ce{Ta2Ni3Te5} monolayer, combining first-principles methods with a semi-empirical approach for computational characterization. As indicated by the phonon dispersion spectrum, the dynamical stability was used to identify the monolayer's active Raman and IR vibrational modes. Our results reveal an anisotropic behavior consistent with the analogous bulk structure. This anisotropy extends to the optical properties, where the optical response is highly sensitive to light polarization due to quantum confinement along the perpendicular basal direction.

\section{Computational details}
The structural, electronic, and vibrational properties were obtained from simulations based on density functional theory (DFT) \cite{Hohenberg_B864_1964, Kohn_1133_1965} methods within the scope of the generalized gradient approximation (GGA)\cite{perdew1996generalized,wu2006more,boese2000new,perdew1991generalized,burke1998derivation} using the exchange-correlation (XC) functional proposed by Perdew--Burke--Ernzerhof (PBE).\cite{perdew1996generalized,perdew1986accurate} Because PBE underestimates the electronic band gap,\cite{Cohen_115123_2008, Crowley_1198_2016} causes self-interaction problems, and gives a poor description of weak interactions,\cite{Rego_415502_2015, Rego_129501_2016, Rego_235422_2017} we employed the hybrid XC functional proposed by Heyd--Scuseria--Ernzerhof (HSE06) \cite{heyd_1187_2004, hummer_115205_2009} to obtain a reasonable correction to the electronic band structure. 

The Kohn--Sham (KS) equations were solved through the projector augmented-wave method (PAW),\cite{blochl1994projector,blochl2005electronic} using the Vienna \textit{Ab Initio} Simulation Package (VASP).\cite{Kresse_13115_1993, Kresse_11169_1996} For all simulations, a total energy convergence criterion of \SI{E-6}{\electronvolt} was employed for the self-consistent cycle. To obtain the equilibrium structures, we optimize the stress tensor and minimize the inter-atomic forces with a plane-wave cutoff energy of \SI{540}{\electronvolt} until the atomic forces on each atom were less than \SI{0.01}{\electronvolt/ \angstrom}. We compute other properties with a lower cutoff energy of \SI{304}{\electronvolt}.

For the integration of the Brillouin zone (BZ), we used a \textbf{k}-mesh of $2\times11\times1$ for the electronic band structure and structural optimization calculations and a denser $4\times22\times1$ \textbf{k}-mesh for the density of states (DOS) calculations. A vacuum distance of \SI{21}{\angstrom} was added along the $z$-direction to eliminate spurious interactions with the structure's mirror images. 

We used the Quantum Espresso (QE) package \cite{giannozzi2009quantum,giannozzi2020quantum} to obtain the phonon dispersion and the phonon density of states of the \ce{Ta2Ni3Te5} monolayer over the entire BZ, using truncated Coulomb interactions, -imposed by the assume$\_$isolated flag,- \cite{sohier2017density}; this method is beneficial to treat two-dimensional systems by avoiding the interaction with repeated images during the phonon calculation. The QERaman code,\cite{hung2024qeraman} interfaced with QE, was then used to determine which of the sixty vibrational modes at the high-symmetry $\Gamma$ point are Raman (R) active. 

Determining the intensity of the R-active modes at several laser energy values can also be inferred, thus allowing a direct comparison with experimental data. These calculations were performed using a \SI{60}{Ry} plane-wave energy cutoff, with a PBE exchange functional embedded in Troullier-Martins (TM) pseudo-potentials (PPs) \cite{troullier1991efficient,engel2001relativistic}. A $2\times10\times1$ \textbf{k}-mesh and a small electronic temperature of \SI{27}{\milli\electronvolt} were added to smoothen the numerical solutions and help reach convergence. It is important to note that the spin-orbit coupling (SOC) has been added self-consistently, only to point out the variation in the band structure. The phonon dispersion, excitonic effects, and Raman spectra were examined without including the SOC.

\begin{figure}[!h]
 \centering
 {\includegraphics[width=1.03\linewidth]{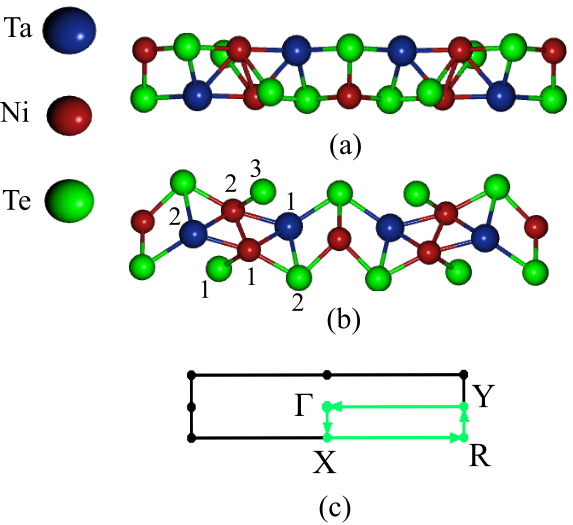}}
  \caption{(Color Online). (a) Top view within the $xy$ plane and (b) side view within the $xz$ plane of a \ce{Ta2Ni3Te5} monolayer with blue, red, and green spheres representing \ce{Ta}, \ce{Ni}, and \ce{Te} atoms, respectively. Some atoms are numbered to refer to different bond lengths. (c) The corresponding rectangular unit cell displaying the $\Gamma$ (0,0,0), X(1/2,0,0), R(1/2,1/2,0), and Y(0,1/2,0) high symmetry points.}
\label{geometry}
\end{figure}

The maximally localized Wannier function Tight Binding (MLWF-TB) method was exploited to describe the electronic states and determine the MLWF-TB Hamiltonian using the Wannier90 code \cite{Arash_685_2008}. The optical properties are then evaluated within the scope of the independent particle approximation (IPA) and through the solution of the Bethe-Salpeter equation (BSE) \cite{Salpeter_1232_1951} using the WanTiBEXOS package.\cite{Dias_108636_2022} It should be noted that the MLWF-TB Hamiltonian was obtained at an HSE06 level, directly from VASP, with a \textbf{k}-mesh of $6\times11\times1$ considering $p$ and $d$ orbital projections for \ce{Ta} and \ce{Ni}, and $s$ and $p$ projections for \ce{Te}. The BSE was solved using the Coulomb truncated 2D potential (V2DT)\cite{Rozzi_205119_2006} with a $7\times32\times1$ \textbf{k}-mesh taking into account the lowest 12 conduction bands and the highest 12 valence bands; also a smearing value of \SI{0.05}{\electronvolt} was applied for the dielectric function computation.

\begin{figure*}[!htbp]
 \centering
 {\includegraphics[width=0.9\linewidth]{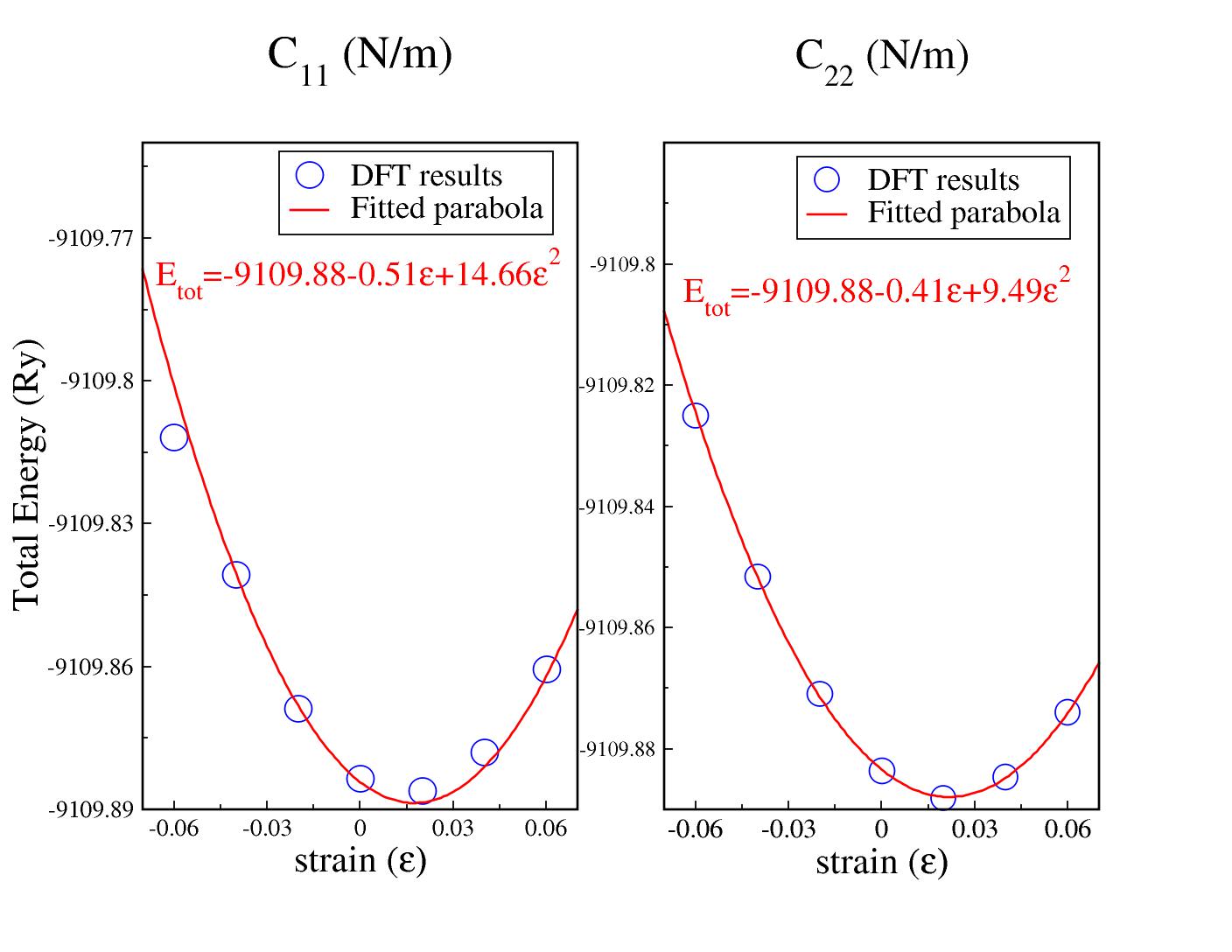}}
  \caption{The elastic constants C$_{11}$ and C$_{22}$, obtained as a response to strain applied along the $x$ and $y$ directions, respectively.}
\label{elastic}
\end{figure*}

\section{Geometry and Structural Properties}
The top and side views of the monolayer, as well as its two-dimensional (2D) Brillouin zone (BZ), are illustrated in Fig.~\ref{geometry}. The \ce{Ta2Ni3Te5} monolayer pertains to the \textit{Pm} space-group with equilibrium lattice constants a=\SI{17.835}{\angstrom} and b=\SI{3.737}{\angstrom}, consistent with the bulk experimental data.\cite{Harrison_4811_2024} The structure has two Ni-Te bond lengths, with $d_{\rm{Ni}_1-\rm{Te}_1}$=\SI{2.58}{\angstrom} and $d_{\rm{Ni}_1-\rm{Te}_3}$=\SI{2.80}{\angstrom}. On the other hand, $d_{\rm{Ni}_1-\rm{Ta}_1}$=\SI{2.68}{\angstrom}, $d_{\rm{Ni}_1-\rm{Ta}_2}$=\SI{2.62}{\angstrom}, $d_{\rm{Ni}_1-\rm{Ni}_2}$=\SI{2.49}{\angstrom} and $d_{\rm{Ta}_1-\rm{Te}_2}$=\SI{2.89}{\angstrom}. The cohesive energy per atom (E$_{coh/atom}$), resulting from the arrangements of the different atoms in the ground state of the \ce{Ta2Ni3Te5} monolayer, is defined via :

\begin{equation*}
E_{coh/atom}=\frac{E_{tot}-4E_{\rm{Ta}}-6E_{\rm{Ni}}-10E_{\rm{Te}}}{20},
\end{equation*}

where E$_{tot}$ is the total energy of the monolayer and E$_{\rm{Ta}}$, E$_{\rm{Te}}$, and E$_{\rm{Ni}}$ are the energies of isolated Ta, Te, and Ni atoms, respectively; the number '20' in the denominator refers to the total number of atoms in the unit cell. Our calculations predict an E$_{coh/atom}$=\SI{-4.492}{\electronvolt}, which, in principle, means that the structure is energetically favorable.

\section{Mechanical Properties}
Studying the structure's response to different strain types lets us obtain the mechanical properties of the \ce{Ta2Ni3Te5} monolayer. Since the unit cell is rectangular, a strain along the $x$ and $y$ directions will cause a change in the lattice parameters along these directions without actually varying the rectangular unit cell. As a consequence, the C$_{11}$ and C$_{22}$ elastic constants can be determined. A third elastic constant, C$_{12}$, results from a biaxial strain within the $xy$ plane that also preserves the rectangular unit cell. 

Fig. \ref{elastic} exhibits the variation of the total energy (E$_{\rm{tot}}$) under compressive and tensile strain ($\varepsilon$) values, ranging from -6$\%$ to 6$\%$ and applied along the $x$ and $y$ directions. The DFT values are fitted to second-degree polynomials in $\varepsilon$, from which C$_{11}$ and C$_{22}$ were determined as $\frac{t}{V_0}\frac{\partial^2 E_{tot}}{\partial\varepsilon^2}$, $t$ being the thickness of the \ce{Ta2Ni3Te5} monolayer. In this way, C$_{11}$ and C$_{22}$ will be vacuum-independent and in units of N/m.

From our calculations C$_{11}$=\SI{98.87}{\newton/\m}, C$_{22}$=\SI{64.05}{\newton/\m}, and C$_{12}$=\SI{21.56}{\newton/\m}. Furthermore, the Poisson ratios $\nu^{x}_{2D}$= C$_{12}$/ C$_{11}$ and $\nu^{y}_{2D}$= C$_{12}$/ C$_{22}$ are 0.22 and 0.34, respectively. These values are vacuum-independent, exhibiting an anisotropy along the $x$ and $y$ directions. They also illustrate the mechanical stability of the structure with C$_{11} >$ 0, C$_{22} >$ 0 and $\nu^{x}_{2D}$ and $\nu^{y}_{2D} <$ 0.5. A positive Poisson's ratio confirms that the material is non-auxetic, contracting along the transverse direction when subjected to tensile forces.

\section{Electronic Band structure and Density of States (DOS)}
 
 \begin{figure}[!h]
 \centering
 {\includegraphics[width=0.95\linewidth]{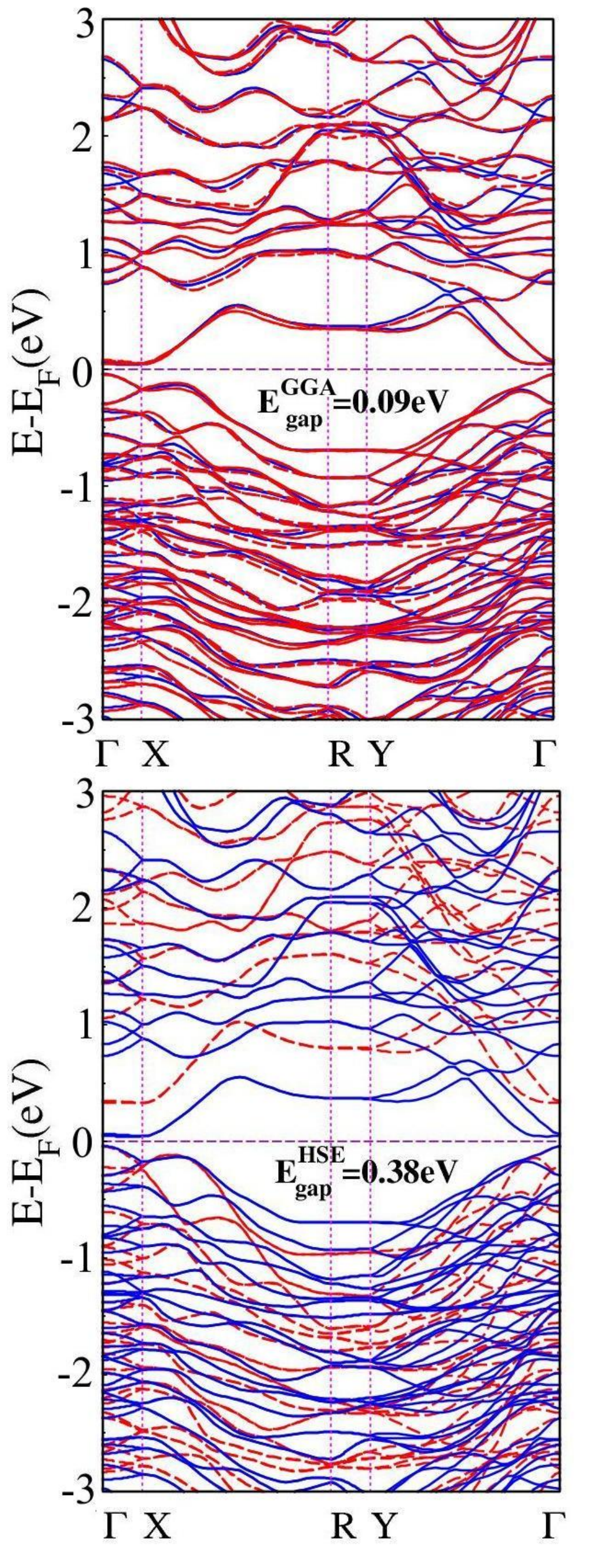}}
  \caption{(Color Online). (Top) Electronic band structure with (in red) and without (in blue) SOC of the \ce{Ta2Ni3Te5} monolayer, with a band gap of \SI{0.09}{\electronvolt}. (Bottom) Comparison between the GGA-PBE (in blue) and the HSE06 (in red) electronic band structures of the \ce{Ta2Ni3Te5} monolayer. The corrected band gap value is \SI{0.38}{\electronvolt}. Here, SOC has not been taken into account.}
\label{bands-DOS}
\end{figure}

 \begin{figure*}[!h]
 \centering
 {\includegraphics[width=0.9\linewidth]{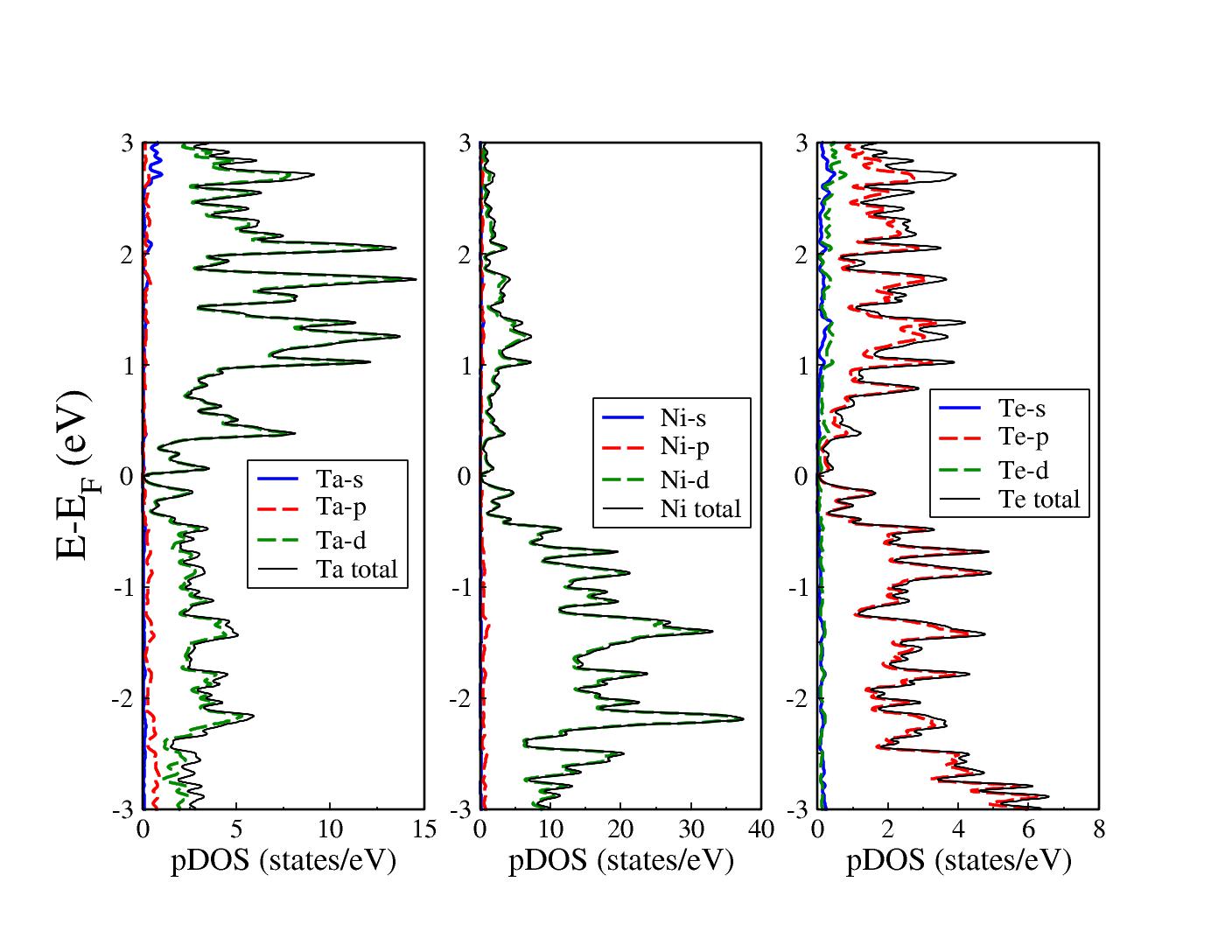}}
  \caption{(Color Online). Partial density of states (pDOS) of the (a) \ce{Ta}, (b) \ce{Ni}, and (c) \ce{Te} atoms in the \ce{Ta2Ni3Te5} monolayer. The valence and conduction bands are mostly composed of the $d$ orbitals of \ce{Ni} and \ce{Ta} and the $p$ orbitals of \ce{Te}. The Fermi level is set at \SI{0}{\electronvolt}.}
\label{pDOS-DOS}
\end{figure*}

\begin{figure*}[htpb!]
 \centering
 {\includegraphics[width=1.0\linewidth]{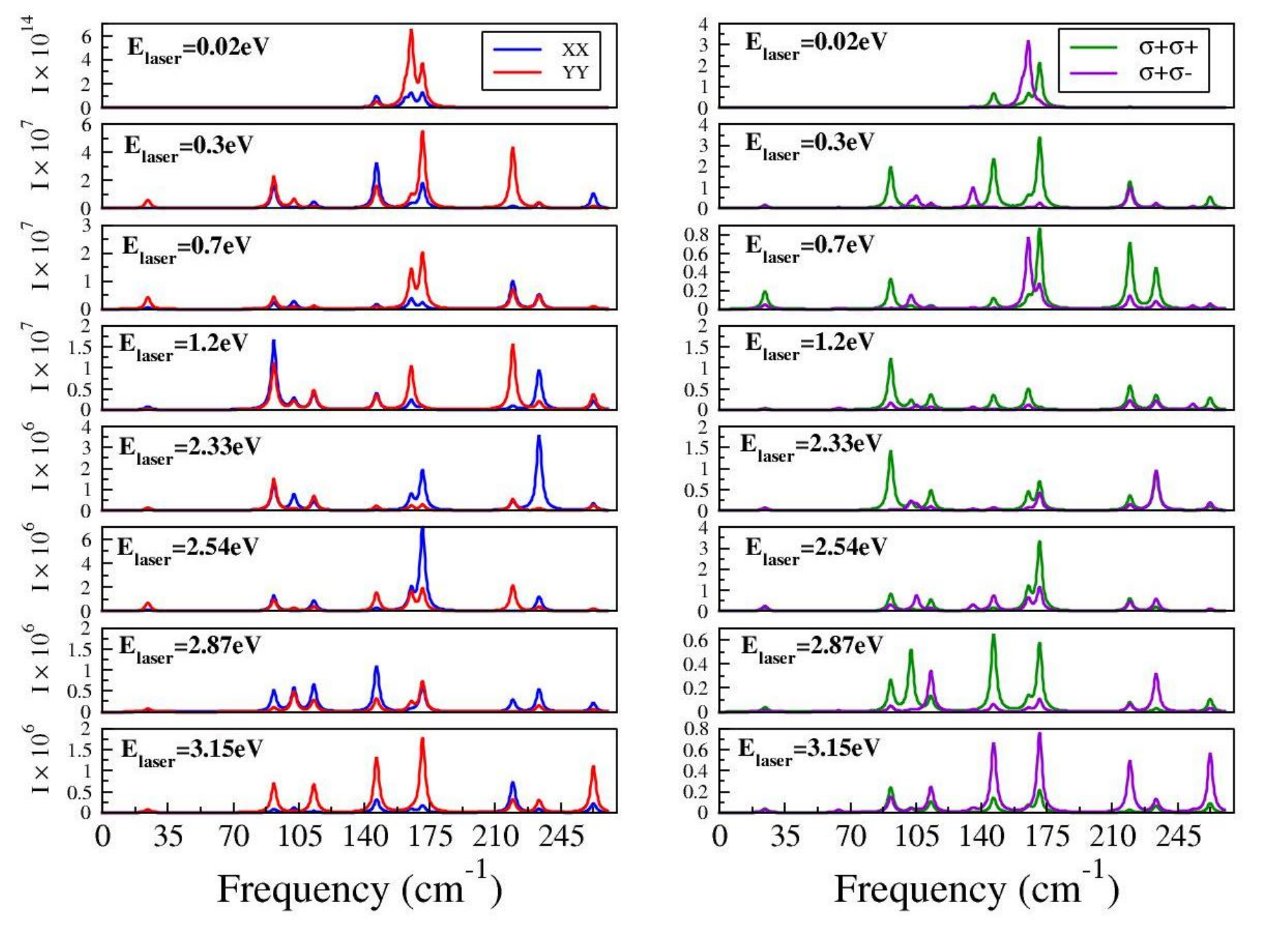}}
  \caption{(Color Online). Raman spectra for a \ce{Ta2Ni3Te5} monolayer for different values of the laser energy (E$_{\rm{laser}}$). Results from the linearly (XX,YY) and circularly polarized light; both helicity-conserving ($\sigma$+$\sigma$+) and helicity-changing ($\sigma$+$\sigma$-) scattering cases are considered. ``I'' refers to the intensity of the observed peaks.}
\label{raman}
\end{figure*}

Our calculations show that the top of the valence band (VBM) occurs at the $\Gamma$ point, while the bottom of the conduction band (CBM) is at a point along the $\Gamma$-X path. Fig.~\ref{bands-DOS} demonstrates that the \ce{Ta2Ni3Te5} monolayer is a semiconductor with an indirect small electronic band gap of \SI{0.09}{\electronvolt}. This result is consistent with Ref. \cite{Guo_87_2022}  (\SI{0.07}{\electronvolt}). The density of states (DOS) presented in Fig.~\ref{pDOS-DOS} shows that within the \SIrange{-3}{3}{\electronvolt} considered energy range, both the valence and the conduction bands are mainly contributed by the $d$ orbitals of \ce{Ni} and \ce{Ta}, and the $p$ orbitals of \ce{Te}. More specifically, the most prominent DOS of the valence bands comes from the $d$ orbitals of \ce{Ni}; in contrast, the corresponding DOS of the conduction bands is primarily influenced by the $d$ orbitals of the \ce{Ta} atoms.

It is essential to mention that the bulk \ce{Ta2Ni3Te5} investigated in Jiang's work,\cite{Jiang_067402_2024} confirms the semiconducting nature of this material with a small band gap of \SI{31}{\milli\electronvolt} using the Projector Augmented Wave (PAW) method \cite{blochl1994projector} with a modified Becke-Johnson functional \cite{koller2011merits,camargo2012performance} to describe the exchange potential. On the other hand, the 2D analog has a band gap three (twelve) times larger on a PBE (HSE06) level.

\section{Phonon dispersion and Raman and IR Active Modes}
Phonon calculations were carried out at the $\Gamma$ high-symmetry point to obtain the modes of vibration of the monolayer. Since the unit cell contains 20 atoms, we expect 60 vibrational modes, the first three of which are acoustic (A) modes and the remaining 57 are optical (O) ones. At the $\Gamma$ point, the modes decompose into:
\begin{equation*}
\Gamma\equiv 20 A_1+10 A_2+ 20 B_1+10 B_2~ \rm{.}   
\end{equation*}
The three acoustic modes belong to the B$_1$, B$_2$, and A$_1$ symmetries, and 10 out of the 20 A$_1$ modes are Raman (R) active.

\begin{figure*}[htpb!]
 \centering
 {\includegraphics[width=1.0\linewidth]{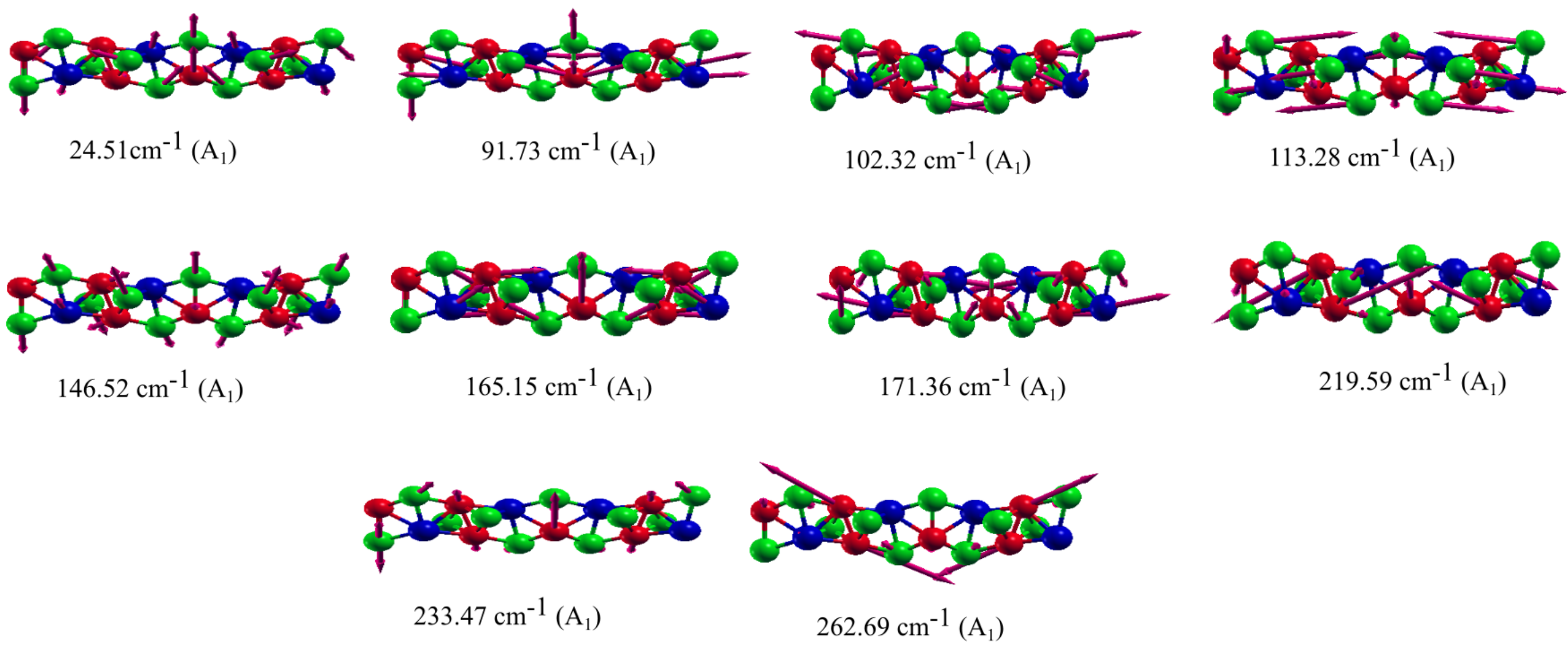}}
  \caption{(Color Online). The ten Raman active modes showing the vibrations of the \ce{Ta}, \ce{Ni}, and \ce{Te} atoms for each mode in the $xz$ plane. The blue, red, and green spheres correspond to the \ce{Ta}, \ce{Ni}, and \ce{Te} atoms, respectively.}
\label{modes}
\end{figure*}

We have estimated the Raman spectra for linearly and circularly polarized light to determine these modes using several laser energy values ranging from the infrared to the ultraviolet regimes. The results are presented in Fig.~\ref{raman}. Our results are based on the assumption that the incident and scattered polarization vectors are parallel and that linearly polarized light propagates in a direction perpendicular to the plane of the \ce{Ta2Ni3Te5} monolayer. Only three peaks can be detected for laser energies (E$_{\rm{laser}}$) as low as \SI{0.02}{\electronvolt}. As E$_{\rm{laser}}$ increases, other peaks emerge, varying in intensity. All in all, we can identify ten such peaks. Focusing on E$_{\rm{laser}}$=\SI{2.33}{\electronvolt}, the peaks appear at \SI{24.50}{\per\cm}, \SI{91.73}{\per\cm}, \SI{102.32}{\per\cm}, \SI{113.28}{\per\cm}, \SI{146.52}{\per\cm}, \SI{165.15}{\per\cm}, \SI{171.36}{\per\cm}, \SI{219.59}{\per\cm}, \SI{233.47}{\per\cm}, and \SI{262.69}{\per\cm}.

\begin{figure*}[htbp!]
 \centering
 {\includegraphics[width=1.0\linewidth]{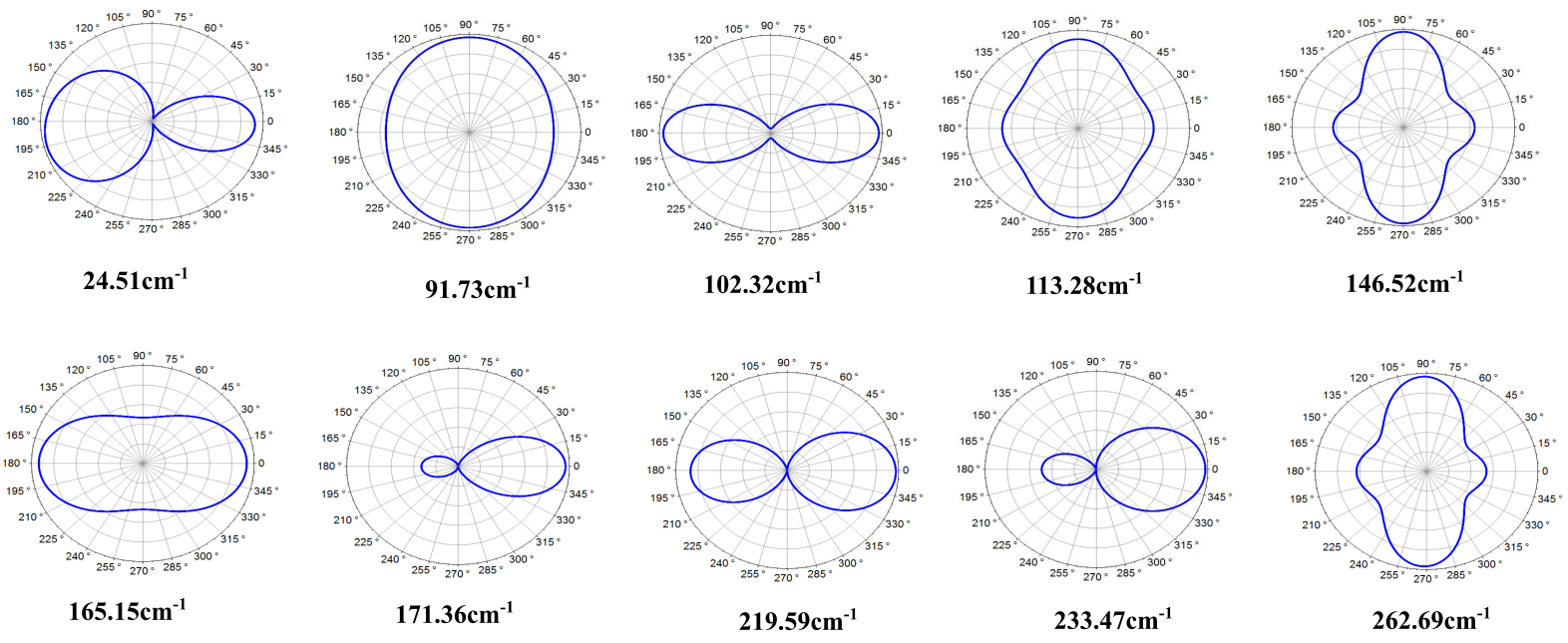}}
  \caption{(Color Online). Based on complex Raman tensors, Raman intensities for circularly polarized light are displayed for the ten active Raman modes of the \ce{Ta2Ni3Te5} monolayer.}
\label{intensity}
\end{figure*}

\begin{figure*}[htbp!]
 \centering
 {\includegraphics[width=1.0\linewidth]{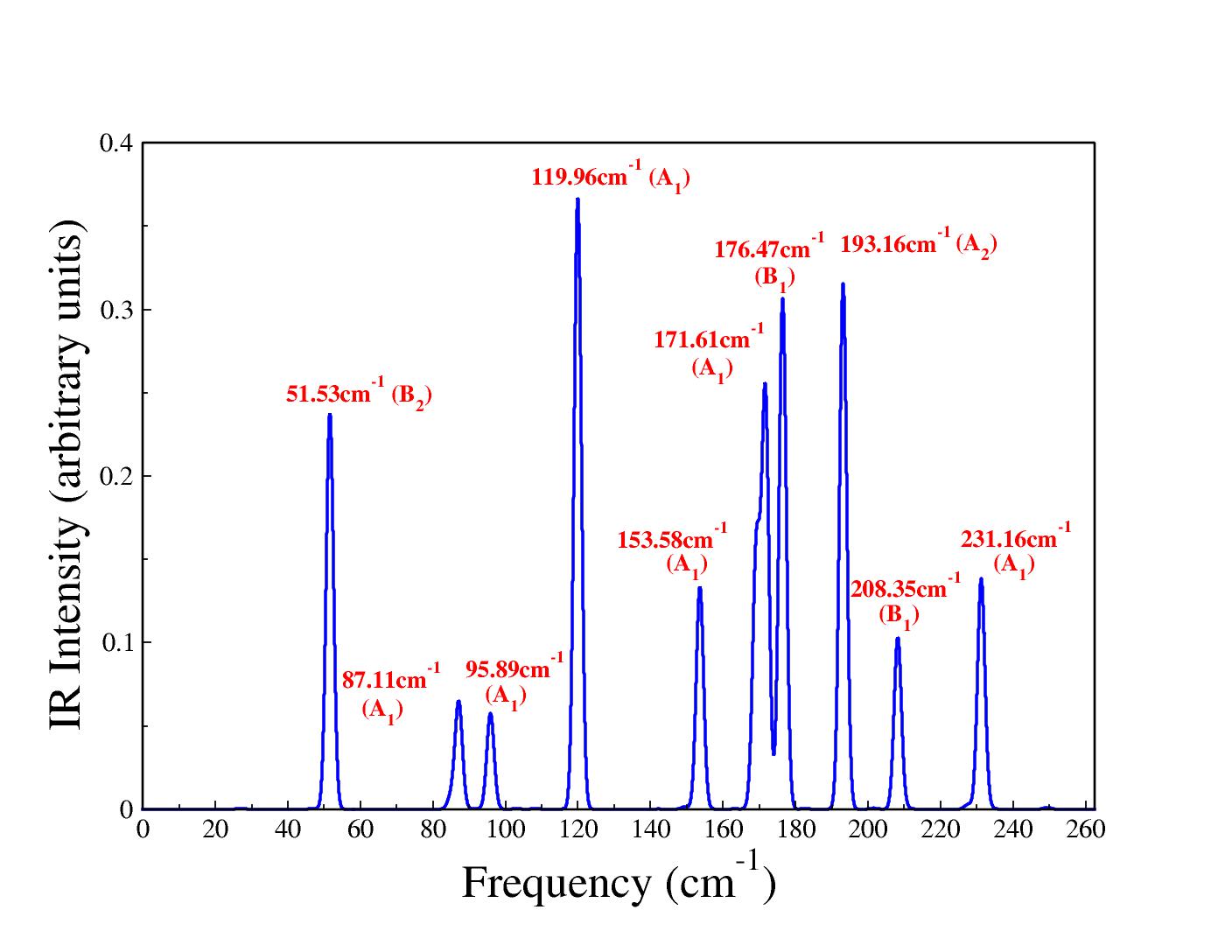}}
  \caption{(Color Online). The IR spectrum of the 2D-\ce{Ta2Ni3Te5} material. The frequencies and symmetry of the modes are also displayed.}
\label{IR-spectrum}
\end{figure*}

The Raman active modes are presented in Fig.~\ref{modes}. We note that most of the vibrational modes do not exhibit the displacement of the \ce{Ni} atoms. On the other hand, mode 5 is mainly contributed by the vibration of the \ce{Te} atoms. Fig.~\ref{intensity} shows the Raman intensity for circularly polarized light for the ten Raman active modes. The intensity of mode 3 reaches its maximum for polarization angles $\theta$=\SI{0}{\degree} or \SI{180}{\degree}, that is, along the $\pm$x-directions. Modes 6, 7, 8, and 9 only present maxima at $\theta$ = \SI{0}{\degree}. In contrast, the intensity maxima of modes 2, 4, 5, and 10 occur when $\theta \sim$ \SI{90}{\degree} and $\theta$=\SI{270}{\degree}, that is, along the $\pm$ y directions, with mode 2 being quasi-isotropic. Mode 1 is an entirely distinct case where the maximum does not occur at \SI{180}{\degree}, but rather at $\theta$=\SI{195.38}{\degree}. 

We have further calculated the IR intensities presented in Fig.~\ref{IR-spectrum}. 
The results indicate that ten of the remaining modes are IR-active. More specifically, peaks of distinct intensities occur at \SI{51.53}{\per\cm}, \SI{87.11}{\per\cm}, \SI{95.89}{\per\cm}, \SI{119.96}{\per\cm}, \SI{153.58}{\per\cm}, \SI{171.61}{\per\cm}, \SI{176.47}{\per\cm}, \SI{193.16}{\per\cm}, \SI{208.35}{\per\cm}, and \SI{231.16}{\per\cm}. Due to the closeness of the \SI{171.61}{\per\cm} and \SI{231.16}{\per\cm} IR frequencies to those of the Raman active ones (\SI{171.36}{\per\cm} and \SI{233.47}{\per\cm}), we speculate that these modes can be both IR+R active. 

\begin{figure*}[!h]
 \centering
 {\includegraphics[width=1.0\linewidth]{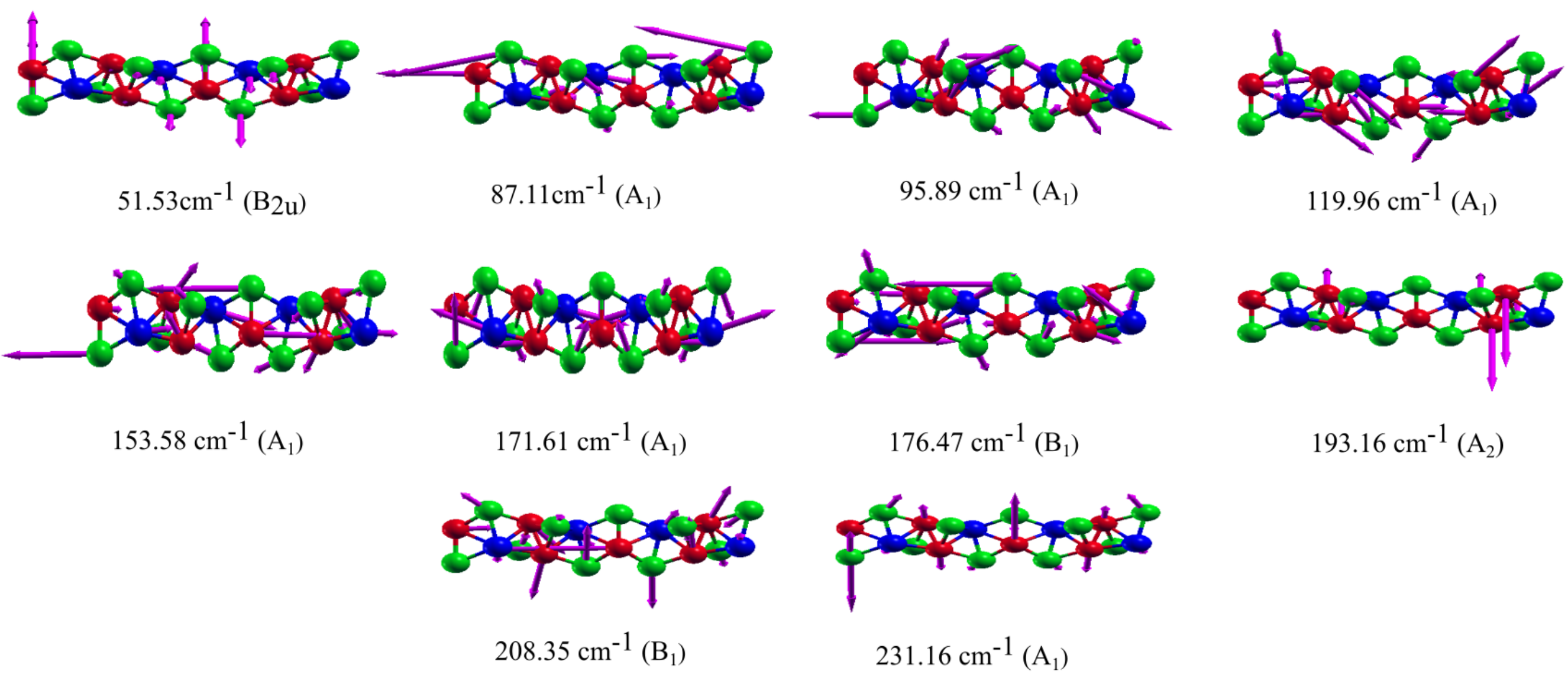}}
  \caption{(Color Online). The identified ten IR active modes, corresponding to the IR spectrum peaks. The blue, red, and green spheres correspond to the \ce{Ta}, \ce{Ni}, and \ce{Te} atoms, respectively. The vibrations are shown in the $xz$ plane.}
\label{IR-vib}
\end{figure*}

The atomic vibrations of the active modes in IR, shown in Fig.~\ref{IR-vib}, belong to the symmetries A$_1$, B$_1$, and B$_{2u}$. Modes 1 and 8 almost vibrate along the $z$ direction and are composed mainly of the vibrations of the Te and Ni atoms. The vibration of the Ta atoms is virtually absent in all IR modes.

Ref.~\cite{Harrison_4811_2024} investigated the Raman active modes of few-layer (FL) \ce{Ta2Ni3Te5} flakes, which were composed of two monolayers. To our knowledge, no work has yet discussed the monolayered structure. In total, fifteen modes have been experimentally observed for the FL \ce{Ta2Ni3Te5} system emerging at \SI{10}{\per\cm}, \SI{28}{\per\cm}, \SI{35}{\per\cm}, \SI{63}{\per\cm}, \SI{86}{\per\cm}, \SI{92}{\per\cm}, \SI{104}{\per\cm}, \SI{116}{\per\cm}, \SI{127}{\per\cm}, \SI{138}{\per\cm}, \SI{153}{\per\cm}, \SI{166}{\per\cm}, \SI{172}{\per\cm}, \SI{204}{\per\cm}, and \SI{228}{\per\cm}. Some of these modes disappeared in the Raman spectrum of the monolayer, such as the \SI{10}{\per\cm}, \SI{35}{\per\cm}, \SI{63}{\per\cm}, and \SI{86}{\per\cm} while others appeared such as the  \SI{262.69}{\per\cm}. 

\begin{figure*}[!h]
 \centering
 {\includegraphics[width=1.0\linewidth]{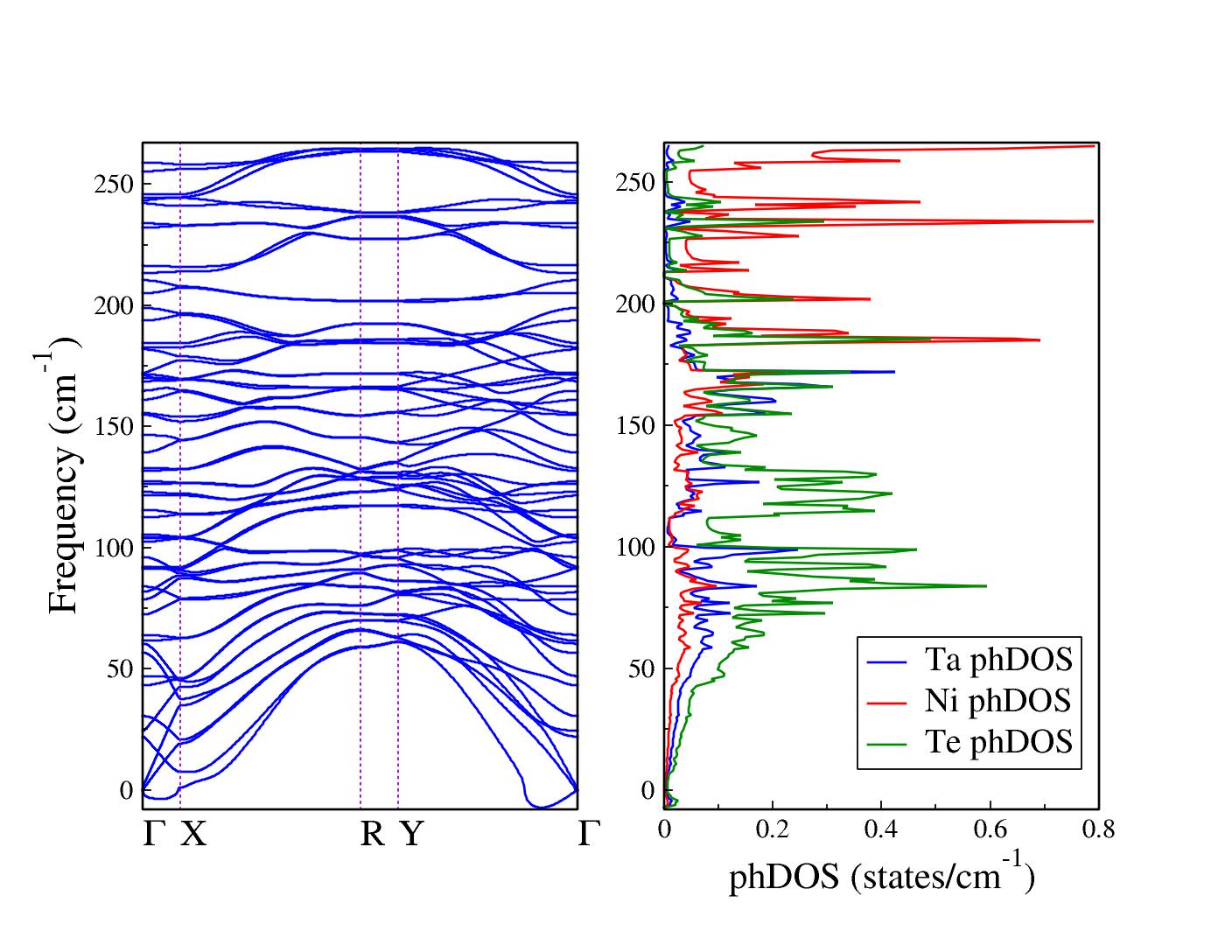}}
  \caption{(Color Online). (Left) Phonon dispersion of the \ce{Ta2Ni3Te5} monolayer along the $\Gamma$-X-R-Y-$\Gamma$ path. (Right) Phonon density of states and the contribution of the Ta, Ni, and Te atoms to the vibrational modes.}
\label{phdisp}
\end{figure*}

To determine the overall contributions of the \ce{Ta}, \ce{Ni}, and \ce{Te} atoms, the phonon density of states (phDOS) is plotted for the different phonon mode regimes, as demonstrated in Fig.~\ref{phdisp}. The phononic band structure, also displayed in Fig.~\ref{phdisp}, shows some slight negative frequencies in the flexural mode, not exceeding \SI{-7.6}{\per\cm}, near the $\Gamma$ point along the Y-$\Gamma$ path. More minor negative frequencies, not ultrapassing \SI{-3.78}{\per\cm}, extend along the $\Gamma$-X path. These frequencies do not indicate a dynamical instability of the monolayer but instead represent numerical inaccuracies due to the diagonalization of the dynamical matrix. The phDOS exhibits the dominant contribution of the \ce{Te} atoms up to $\sim$ \SI{170}{\per\cm}. For modes of higher frequencies, it is evident that the \ce{Ni} atoms contribute the most.

\section{Excitonic and Optical Properties}

 \begin{figure}[!h]
 
 {\includegraphics[width=1.0\linewidth]{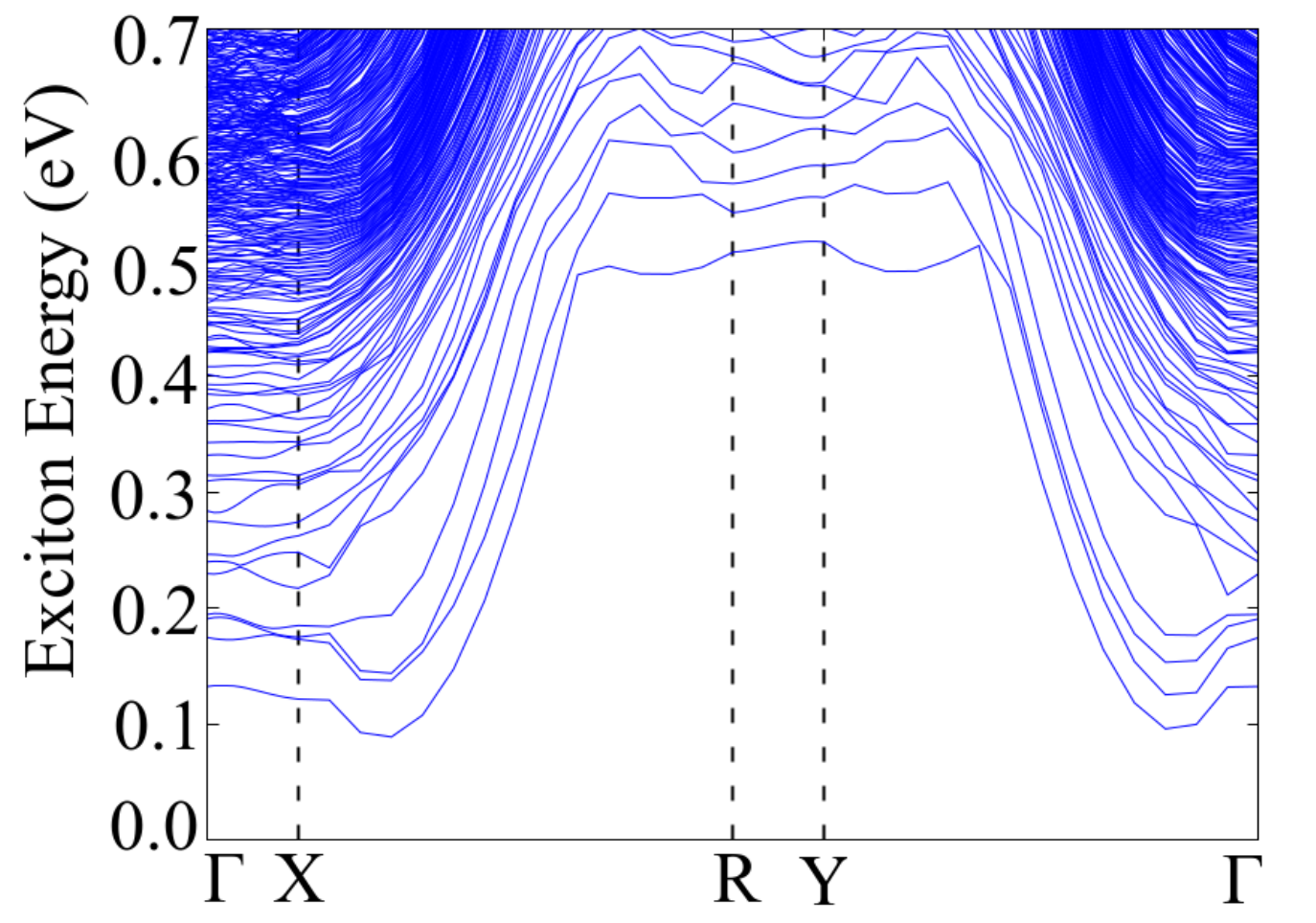}}
  \caption{(Color Online). MLWF-TB HSE06 \ce{Ta2Ni3Te5} monolayer excitonic band structure.}
\label{exc_bands}
\end{figure}

From the excitonic band structure, shown in Fig.~\ref{exc_bands}, we can observe that the exciton ground state is indirect, with an exciton momentum between the $X$ and $R$ high symmetry points and an exciton binding energy of \SI{287}{\milli\electronvolt} which lies in the expected range for 2D materials.\cite{Dias_3265_2021} The direct exciton ground state, located at $\Gamma$, has a \SI{0.13}{\electronvolt} gap and corresponds to the optical band gap. The presence of an indirect exciton ground state suggests the possibility of phonon-assisted optical transitions with excitation energies lower than the optical band gap; in fact, the difference between the exciton ground state and optical band gap is very small, which makes it difficult to identify both peaks in the optical spectrum.

\begin{figure}[!h] {\includegraphics[width=0.9\linewidth]{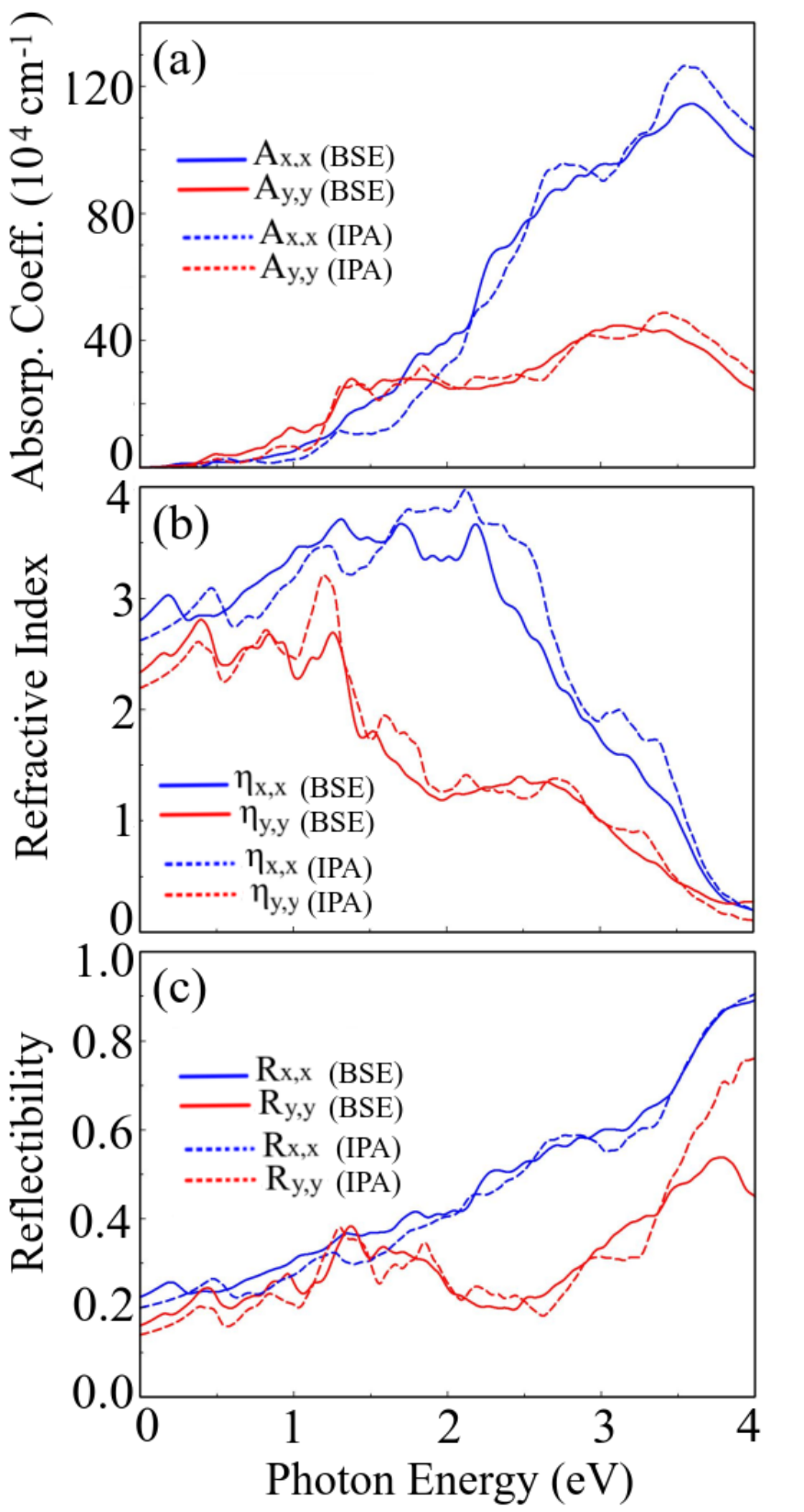}}
  \caption{(Color Online). MLWF-TB HSE06 \ce{Ta2NiTe5} monolayer optical properties: (a) Absorption coefficient, (b) refractive index, and (c) reflectivity, all determined at BSE (solid curves) and IPA (dashed curves) levels, considering linear $\hat{x}$ (blue curves) and $\hat{y}$ (red curves) polarizations.}
\label{optics}
\end{figure}

The linear optical response of the \ce{Ta2Ni3Te5} monolayer is shown in Fig.~\ref{optics}, with (BSE - solid curves) and without (IPA - dashed curves) excitonic effects for linearly polarized light along the $\hat{x}$ (blue curves) and $\hat{y}$ directions (red curves). From the absorption spectrum, shown in Fig.~\ref{optics}(a), we can observe that for lower optical excitations in the infrared and at the beginning of the visible spectrum (around \SI{1.5}{\electronvolt}), the optical response is very similar, independent of the light polarization. Conversely, we can see that, for higher excitations, the system absorbs with a higher intensity for the $\hat{x}$ polarization case, showing a significant optical anisotropy in the visible and ultraviolet regions. The excitonic effects result in a slight red shift in the absorption spectrum, yet these quasi-particle effects do not considerably change the spectrum shape and intensities.

The refractive index and reflectivity are shown in Figs.~\ref{optics} (b) and (c), respectively. The refractive index is higher for polarization along the $\hat{x}$ direction. It registers the highest value in the visible region, decreasing as the photon energy moves to the ultraviolet region. Although the excitonic effects are evident, the changes are inconsiderable. Regarding the reflectivity, the opposite happens, as this factor is enhanced as the photon excitation energy increases. For photon energies higher than \SI{2}{\electronvolt}, the optical anisotropy becomes more evident, showing a quasi-total reflectivity for photons closer to the ultra-violet region for light polarized along the $\hat{x}$ direction. For a polarization along the $\hat{y}$ direction and at BSE (IPA) level, the reflectivity at \SI{4}{\electronvolt} is \SI{40(80)}{\percent}. The comparison between BSE and IPA optical responses is vital as the excitonic effects are susceptible to the substrate where the monolayer could be placed. The higher the dielectric constant of the substrate is, the lower the excitonic effects \cite{Dias_3265_2021}. 


Besides, we have observed an interesting behavior when photon energy increases and is in the ultraviolet (UV) region; the absorption and refractive index tend to decrease while reflectivity increases, indicating that the material minimally absorbs ultraviolet radiation and has a high potential for reflection. This comportment is particularly relevant for materials used to fabricate ultraviolet-blocking devices.

\section{Conclusions}
The Raman spectra, the excitonic effect, and the dynamical and mechanical stabilities of the novel \ce{Ta2Ni3Te5} monolayer have been investigated. The monolayer is an indirect small band gap semiconductor of \SI{0.09}{\electronvolt} (\SI{0.38}{\electronvolt}) band gap value within the PBE (HSE06) approximation. It exhibits anisotropic properties, registering different mechanical responses to strain when applied along the $x$ and $y$ directions. It is also dynamically stable, except for minor negative frequencies near the $\Gamma$ point in the corresponding phonon dispersion spectrum, primarily due to numerical inaccuracies in the dynamical matrix diagonalization process. Its phonon DoS illustrates the dominant contributions of the Te and Ni atoms to the vibrational modes of the monolayer.

The Raman spectrum exhibits an increase in the number of peaks upon increasing the laser energy values. We have identified ten peaks corresponding to this monolayer's ten Raman active modes for a typical laser energy of \SI{2.33}{\electronvolt}. We have further determined ten infrared modes characterized by the predominant vibrations of the \ce{Te} and \ce{Ni} atoms.

The excitonic band structure for this monolayer possesses a binding energy within the expected range for 2D materials and an indirect exciton ground state, which is a signature of the possibility of the occurrence of phonon-assisted optical transitions with excitation energies smaller than the optical bandgap. Additionally, the system is highly anisotropic, causing an optical response dependent on the incident light polarization. Moreover, the system shows a higher reflectivity for light polarized along the $\hat{x}$ direction in the visible and ultraviolet regions, making the \ce{Ta2Ni3Te5} monolayer appropriate as a polarizing filter. 

\begin{acknowledgement}
The authors are grateful for the computational resources provided by Centro Nacional de Processamento de Alto Desempenho in São Paulo-CENAPAD-SP (proj 897, 909 and 950) and the Lobo Carneiro HPC (NACAD) at the Federal University of Rio de Janeiro (UFRJ) (proj 133 and 135). A.C.D. also thanks the financial support from National Council for Scientific and Technological Development (CNPq, grant number $408144/2022-0$ and $305174/2023-1$), Federal District Research Support Foundation (FAPDF, grants number $00193-00001817/2023-43$ and $00193-00002073/2023-84$) and PDPG-FAPDF-CAPES Centro-Oeste (grant $00193-00000867/2024-94$). RMT and DSG acknowledge support from CNPq and the Center for Computational Engineering and Sciences at Unicamp, FAPESP/CEPID Grant ($2013/08293-7$). Elie A. Moujaes would like to thank the financial support of the Brazilian National Council for Scientific and Technological Development CNPq, grant number $315324/2023-6$.

\end{acknowledgement}

\clearpage
\bibliography{zboxref/ref}
\end{document}